# Interactive Art To Go

**Ichiroh Kanaya**
Graduate School of Engineering,
Osaka University
Suita, Osaka, Japan
kanaya@pineapple.cc

**Masataka Imura**
Graduate School of
Engineering Science,
Osaka University
Toyonaka, Osaka, Japan
imura@bpe.es.osaka-u.ac.jp

**Mayuko Kanazawa**
MayuArt.com
Kobe, Japan
kanazawa@mayuart.com

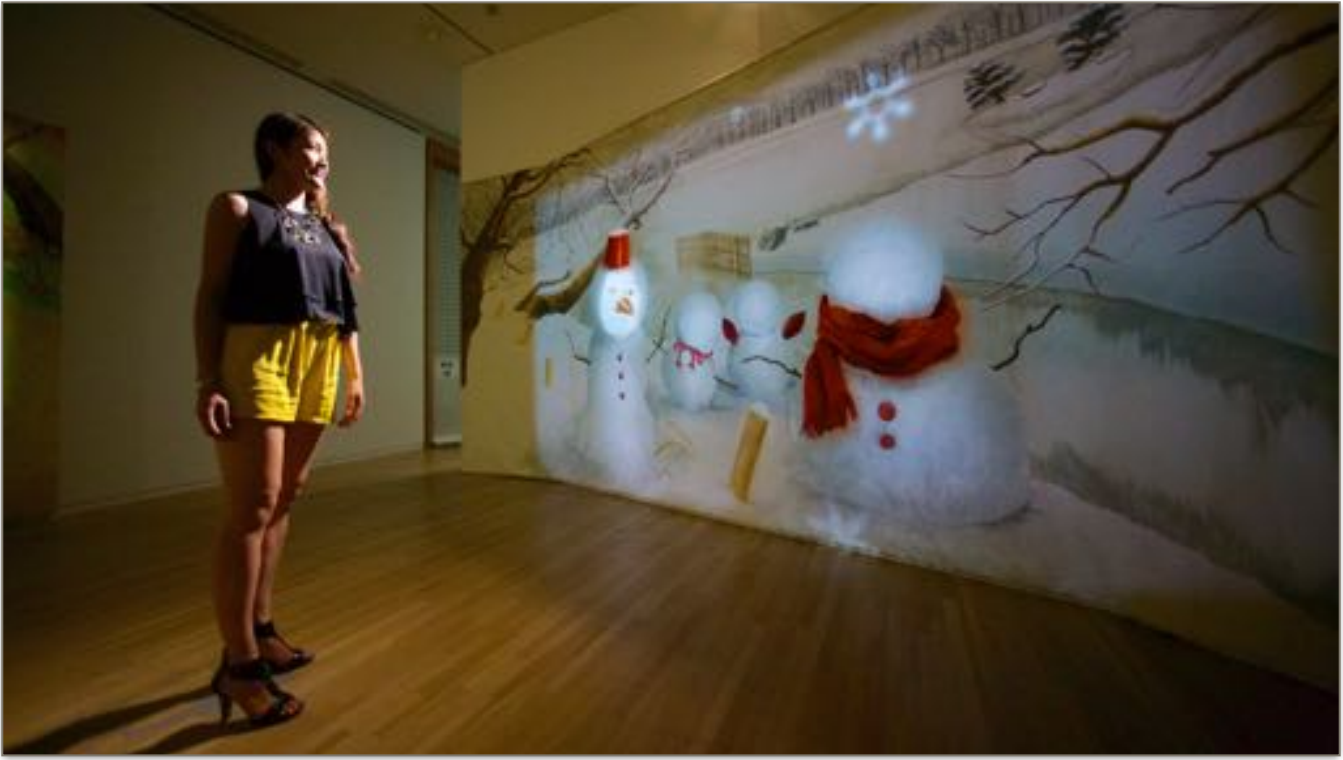

**Figure 1: Yukinko, a new media art created by the authors, at Museum of Contemporary Art Tokyo.**

## ABSTRACT

Traditional artworks like paintings, photographs, or films can be reproduced by conventional media like printing or video. This makes visitors of museums possible to purchase postcards, posters, books, and DVDs of pictures and/or movies shown at the exhibition. However, newly developing arts so called interactive art, or new media art, has not been able to be reproduced due to limitation of functionalities of the conventional media. In this article, the authors report a novel approach of sharing such interactive art outside the exhibition, so that the visitors of the museum can take a copy to home, and even share it with non-visitors. The authors build up their new projector-and-camera (ProCam) based interactive artwork for exhibition at Museum of Contemporary Art Tokyo (MOT) by using Apple's iPhone. The exactly same software driving this artwork was downloadable from Apple's App Store — thus all visitors or even non-visitors could enjoy the same experiment at home or wherever they like.

## ACM Classification Keywords

H.5.1 Animation, J.5 Arts and humanities.

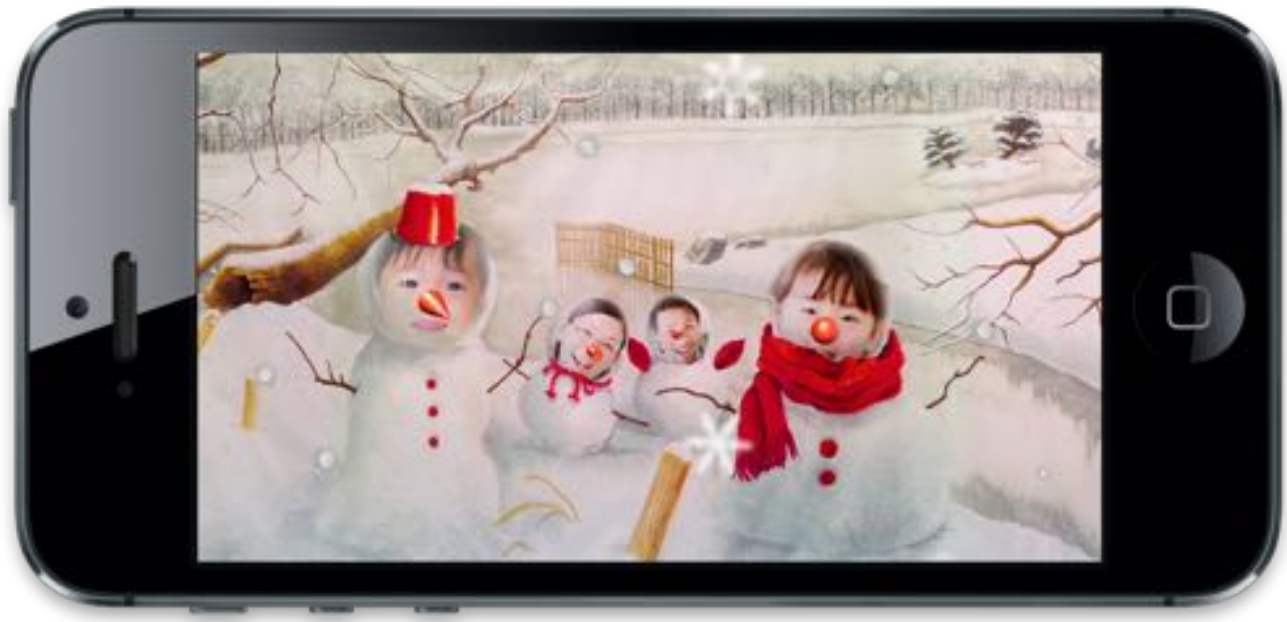

Figure 2: The Yukinko App for iPhone.

**INTRODUCTION**

Traditional artworks, e.g., still images and moving pictures, can be reproduced by conventional media like printing or video disks. Museum shops have been selling replicas of such artworks of exhibitions for visitors so that they can bring those ones, e.g., postcards, posters, DVDs, home for their memories and also sharing their experiences with non-visitors. However, newly developing arts so called interactive arts or (new) media arts have not been able to be photocopied due to limitation of functionalities of conventional media. Such media arts needs computers and often projectors and/or cameras. This was an artists' dilemma — photo and video are not capable to carry the artists' idea, still they are the only ways to share the artists' idea with the world [1–4]. (One giant exception is a computer game in its late golden age. People could play almost the same game of video arcade games by some home entertainment systems like Sony's PlayStation [5]).

On the other hand, many people nowadays bring ultimate computers, equipped with large screens and cameras, almost all day. One possible idea to make an interactive art portable is to port the entire system of the artworks to smartphones, with which many people are bringing. The authors, however, chosen the other extreme: the authors build up the entire system of their artwork, named Yukinko (see Figure 1), by using Apple's iPhone, and pushed the exact same program they created to Apple's App Store so that everyone can download the program to their iPhones (see Figure 2).

Yukinko is a new interactive art that people can interact with a large oil-panted canvas. The canvas has 5m wide and 2.5m height, and is painted with winter countryside scene. Four snow figures without faces are painted on the canvas. Animation of falling snow flakes are projected on the canvas.

When visitors come in front of the canvas, their faces are tracked, captured, and superimposed on the canvas so that they will see their faces as if they become snow figures. Figure 3 demonstrates face detection technology of Yukinko based on OpenCV.

The Yukinko has been exhibited at Museum of Contemporary Art Tokyo (MOT) from 12th July 2014 and scheduled to finish on 31st August. As of two days after the opening, more than 3,000 people aged from 3 to over 70 visited the museum and enjoyed the exhibition. The full report of the exhibition including metrics and questionnaire will be presented at the conference.

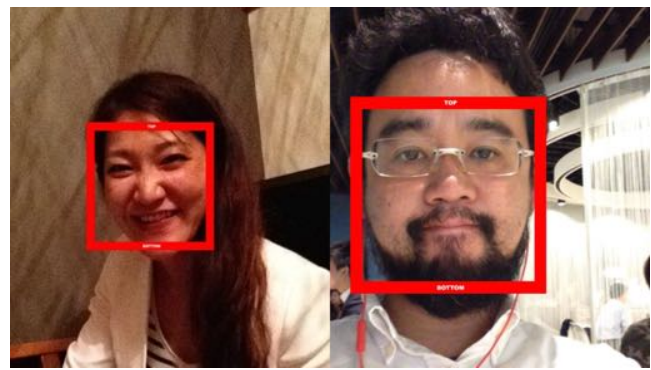

Figure 3: Demonstration of Yukinko's face detector.

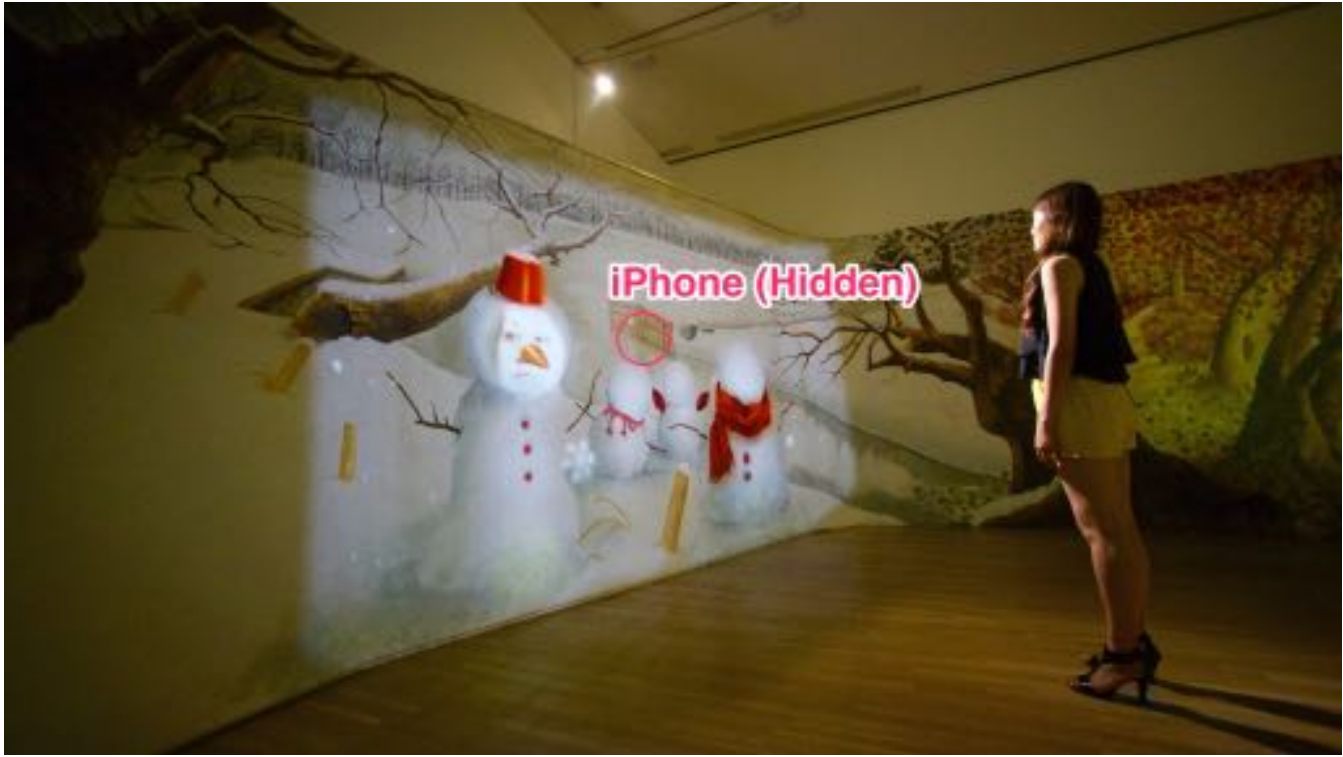

**Figure 4: Set up of Yukinko at Museum of Contemporary Art Tokyo.**

The authors report how they created the exhibition version and mobile application version of Yukinko simultaneously and how they set up the system. The knowledge discovered during setting-up and running period is also reported in this article.

## YUKINKO

Yukinko is, as described quickly in Section 1, a new interactive art that people can interact with an oil-painted canvas. There are four face-less snow figures and background image of snowing country side on the front surface of the canvas. A projector overlays falling snow flakes on top of the front surface of the canvas.

An iPhone 5S (1.3GHz Apple A7, equivalent to ARMv8-A dual core CPU plus PowerVR G6430 GPU, with 1GB RAM) running iOS 7.1.2 is set right behind the canvas. The rear camera of the iPhone 5S (1920 x 1080pix, 30fps, F2.2) captures the scene of the exhibition room through a pin hall of the canvas. Figure 3 and 4 illustrate how the iPhone is installed.

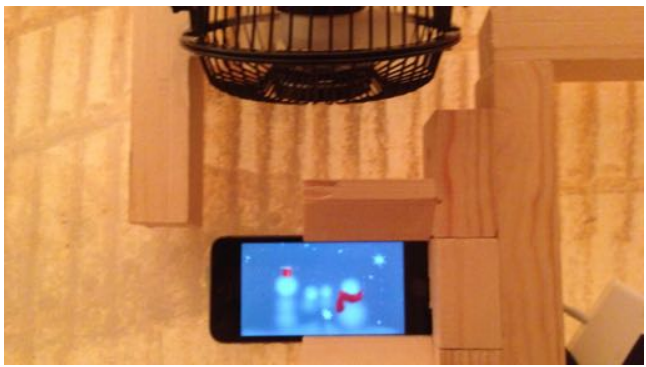

**Figure 5: The behind-the-canvas iPhone with cooling fan.**

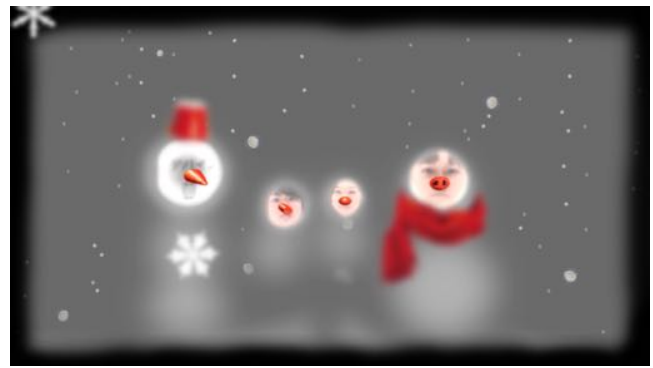

**Figure 6: Synthesized image of animating snow flakes, captured faces, and pre-rendered background animation.**

To avoid heat-up of the iPhone, cooling fan is installed above the iPhone. Thanks to the fan, the temperature of the iPhone has been kept around 25C while it gets over 40C without fan.

A face detecting technology based on OpenCV 2 is used to capture up to four visitors' faces. Captured images of the faces and pre-rendered background animation are superimposed by Apple's Core Animation technology. (Core Animation is an API for layering 2D animations by using OpenGL ES texture buffers.) Animating snow flakes are rendered in real time and also superimposed by Core Animation (see Figure 5).

The rendered frame is sent via Apple's Lightning-VGA adaptor and 10m-long VGA cable to a projector which has 1280x800pix resolution at 60Hz. A 12W battery charger is also installed for stable power supply.

As the program runs on a smartphone, an administrator of the museum easily shut down (put to sleep) and start up (wake up) the system. There are, however, drawbacks of easy sleep/wake interface from the view point of programming of interactive system. Though the conventional software often rely on a set-up-then-run-then-shut-down model, the smartphone programs are on run-sleep-run cycle. Thus the program is designed to detect statuses of (1) the iPhone is about to enter sleep mode, and (2) it is just awake from sleep mode using Apple's Cocoa Touch API, and does necessary cleaning-up process right before sleeping and does necessary starting-up process right after waking up.

The executable of the program, named Yukinko App, can be downloaded from Apple's App Store, and all the source code will be available on GitHub on and after the conference.

**CONCLUDING REMARKS AND FUTURE WORKS**

The authors demonstrated a novel approach to share an interactive art by using a smartphone. The same program works for not only exhibition at museum but also general users. This approach makes the artwork very easy to share all over the world while keeping the effort of making artwork in the museum minimum. It also bring a good side effect that museum administrators can easily turn on and off the system — just pressing sleep button of the iPhone works perfectly. Another good side effect of using smartphone is its modularity. When the system is crashed by some reason, the core module (iPhone) can be easily replaced. All we need is unplugging Lightning cable and audio cable from the original iPhone, replacing it, and plugging those two cables.

Yet another good side effect of using smartphone is that, say, it is a phone. A built-in internet connectivity without WiFi and long-life battery can be used for remote diagnosis even the electric power is cut.

The Yukinko will have been exhibited for about 50 days at the MOT, which is the largest museum for modern art in Japan and expecting around 100,000 visitors. The authors will report how well the system will have run, or survive, this period with all analytics, logs, and questionnaire, as well as demonstrate the iPhone app at the conference.

The authors is also showing three other interactive arts simultaneously at the same room, and two of them are using Kinect 3D sensors, PCs, and projectors with large canvases. Porting such systems to smartphones will be a future work as smartphone developers will soon integrate 3D sensors like Kinect to their smartphone.

**ACKNOWLEDGEMENT**

The authors address special thanks to Mr. Yuta Ideguchi for developing iPhone program and to Dr. Eri Itoh for the photography. The authors also address Mr. Shinji Tanaka, Super Factory, and Museum of Contemporary Art Tokyo for all arrangement of this artwork.